\documentclass[a4paper, 11pt, onecolumn, notitlepage, oneside, reqno]{amsart}
\usepackage{a4wide}
\usepackage[english]{babel}
\usepackage{csquotes}
\usepackage{amsmath}
\usepackage{amsthm}
\usepackage{latexsym}
\usepackage{amssymb}
\usepackage{mathrsfs}
\usepackage{epic}
\usepackage{epsfig}
\usepackage[dvipsnames]{xcolor}
\usepackage{dsfont}
\usepackage[colorlinks = true]{hyperref}
\usepackage{comment}
\usepackage[all]{xy}
\usepackage{enumitem}
\usepackage{eucal}
\usepackage{calc}
\usepackage{multicol}
\usepackage[all]{xy}
\usepackage[final]{showlabels}
\usepackage{tensor}

\usepackage{xcolor}

\usepackage{color}

\newcounter{mnotecount}[subsection]

\newcommand{\hnabla}{\overset{h}{\nabla}}
\newcommand{\vnabla}{\overset{v}{\nabla}}

\newtheorem{theorem}{Theorem}[section]
\newtheorem*{theorem*}{Theorem}

\newtheorem{lemma}[theorem]{Lemma}

\numberwithin{equation}{section}

\begin{document}

\title{Propagation of polarized gravitational waves}

    \author{Lars Andersson}
    \email{lars.andersson@aei.mpg.de}
	\address{Max Planck Institute for Gravitational Physics (Albert Einstein Institute), Am M\"uhlenberg 1, D-14476 Potsdam, Germany}

    \author{J\'er\'emie Joudioux}
    \email{jeremie.joudioux@aei.mpg.de}
	\address{Max Planck Institute for Gravitational Physics (Albert Einstein Institute), Am M\"uhlenberg 1, D-14476 Potsdam, Germany}
	
    \author{Marius A. Oancea}
	\email{marius.oancea@aei.mpg.de}
	\address{Max Planck Institute for Gravitational Physics (Albert Einstein Institute), Am M\"uhlenberg 1, D-14476 Potsdam, Germany}
	
	\author{Ayush Raj}
	\email{ayush.raj@iitb.ac.in}
	\address{Department of Physics, Indian Institute of Technology, Powai, Mumbai-400076, India}

\begin{abstract} The propagation of high-frequency gravitational waves can be analyzed using the geometrical optics approximation. In the case of large but finite frequencies, the geometrical optics approximation is no longer accurate and polarization-dependent corrections at first order in wavelength modify the propagation of gravitational waves, via a spin-orbit coupling mechanism. We present a covariant derivation from first principles of effective ray equations describing the propagation of polarized gravitational waves, up to first-order terms in wavelength, on arbitrary spacetime backgrounds. The effective ray equations describe a gravitational spin Hall effect for gravitational waves, and are of the same form as those describing the gravitational spin Hall effect of light, derived from Maxwell’s equations.
\end{abstract}
	
\maketitle

\section*{Introduction}

The advent of gravitational wave observations brings a new range of phenomena related to the dynamics of the gravitational field to our attention. Gravitational waves propagate over cosmological distances and carry, in addition to information about their sources,  imprints of cosmological expansion and inhomogeneities in the universe. The fact that the important sources of gravitational waves emit in a very broad  range of wavelengths \cite{2005NJPh....7..204F} makes it essential to include effects beyond geometrical optics on their propagation, when considering lensing of gravitational waves \cite{2020arXiv200812814M, PhysRevD.101.044041}. 

Spin-orbit couplings play an essential role when analyzing the propagation of spinning particles and fields in inhomogeneous media beyond the geometrical optics and test particle limit \cite{SOI_review}. For the spin-1 Maxwell field, the spin Hall effect of light has been verified experimentally. When the wavelength is small in comparison with the inhomogeneity scale of the media, a wave packet undergoes a polarization-dependent deviation of the propagation of light beams from the path predicted by geometrical optics \cite{SHEL_experiment,Bliokh2008}. This can be viewed as a manifestation of spin-orbit coupling via the Berry curvature. In general relativity, the dynamics of spinning particles is described by the Mathisson-Papapetrou-Dixon equations \cite{mathisson2010republication,papapetrou1951spinning,tulczyjew1959motion,dixon1964covariant,dixon2015new}, with a suitable closure relation, the so-called spin-supplementary condition. 

Polarization-dependent effects for the propagation of Maxwell fields in curved spacetimes have been discussed previously in Refs. \cite{Harte2018,Frolov,shoom2020,Frolov:2020aa,SHE_QM1}. A detailed review and further references can be found in Ref. \cite{GSHE_review}. Recently, a covariant derivation of the gravitational spin Hall effect of light, based on first principles, has been given in Ref. \cite{GSHE2020}. Similarly, the effective ray equations for massive spin-$\tfrac{1}{2}$ Dirac fields, beyond the geometrical optics limit, have been discussed in Refs. \cite{audretsch,rudiger,Oanceathesis}. The spin-2 nature of the gravitational field leads one to expect that corrections to geometrical optics, involving the Berry curvature, will be relevant also for gravitational waves \cite{SHE_GW,covariantSpinoptics}. 

In this paper, we present the first covariant analysis of the spin Hall effect for gravitational waves. Following the strategy developed in Ref. \cite{GSHE2020} for the Maxwell field, as well as the general theory given in Ref. \cite{Littlejohn1991}, we provide a derivation from first principles of effective ray equations describing the propagation of gravitational waves, up to first-order terms in wavelength, on arbitrary spacetime backgrounds. The equations of motions are obtained through a higher-order geometrical optics approximation using a Wentzel-Kramers-Brillouin (WKB) ansatz. The dynamics of the polarization is described in terms of the Berry connection, and terms of first order in wavelength in the effective ray equations involve the Berry curvature, manifesting the spin nature of the gravitational field. These corrections to the standard trajectories of geometrical optics, the null geodesics, may be termed as the spin Hall effect of gravitational waves \cite{SHE_GW}. It can be shown that the equations of motion are of the same nature as the Mathisson-Papapetrou-Dixon equations for massless spinning particles \cite{HarteOancea, Oanceathesis}, completed by the Corinaldesi-Papapetrou spin supplementary condition (see \cite[Section 3.2.1]{costa2015center}). Our treatment is covariant and applicable to arbitrary curved spacetimes, in contrast to previous work present in the literature. For example, the derivation of the spin Hall effect for gravitational waves given in Ref. \cite{SHE_GW} is not explicitly covariant, and it is limited to propagation in static spacetimes in the weak-field limit. Our derivation of the effect is obtained from the classical field theory of linearized gravity, in contrast with Ref. \cite{SHE_GW} where the author argues that the effect is quantum in nature. Another derivation of a spin Hall effect for gravitational waves was proposed in Ref. \cite{covariantSpinoptics}. While this approach is manifestly covariant, it is limited to stationary spacetimes.

Our starting point is the classical field theory of linearized gravity, governed by a truncated form of the Einstein-Hilbert Lagrangian. A metric perturbation in the form of a WKB ansatz is inserted in the action for linearized gravity, and the resulting expression is truncated after the first order in the inverse of the frequency. This provides a Lagrangian representing the WKB approximation of the linearized gravity field theory. The corresponding Euler-Lagrange equations, with Lorenz gauge imposed, provide the dispersion relation and the transport equation for the amplitude. The dispersion relation is used to define a Hamiltonian for the effective ray equations.

The paper is organized as follows. Sec. \ref{sec:Maxwell_WKB} contains the general setup. The basic equations for linearized gravity are presented in Sec. \ref{sec:linearised}, the gauge choice is discussed in Sec. \ref{sec:lorenzgauge}, and the WKB Ansatz is introduced in Sec. \ref{sec:wkbansatz}. The WKB approximation of the action is made in Sec. \ref{sec:Higher_GO}, and it is shown how the well-known results of geometrical optics can be obtained from the corresponding Euler-Lagrange equations. In Sec. \ref{sec:polarization basis} we discuss the dynamics of the polarization tensor in terms of the Berry connection. The effective dispersion relation is derived in Sec. \ref{sec:effectivedispersion}. Finally, the effective ray equations are discussed in Sec. \ref{sec:effectiveray}. Appendix \ref{app:symbol} contains a discussion on some algebraic property of the symbol. Appendix \ref{app:linearization} presents a self-contained derivation of the equation of linearized gravity. Appendix \ref{app:gauge} contains a basic discussion of the Lorenz gauge.

\section*{Notations and conventions}
We consider an arbitrary smooth Lorentzian manifold $(M, g_{\mu \nu})$, where the metric tensor $g_{\mu \nu}$ has signature $(-\,+\,+\,+)$. The absolute value of the metric determinant is denoted as $g = |\det g_{\mu \nu}|$. The phase space is defined as the cotangent bundle $T^*M$, and phase space points are denoted as $(x, p)$. The Einstein summation convention is assumed. Greek indices represent space-time indices, and run from $0$ to $3$. Latin indices, $(a, b, c, ...)$, represent tetrad indices and run from $0$ to $3$. We  adopt the curvature conventions of \cite{hawking1973}.

\section{The Einstein field equations and linearized gravity} \label{sec:Maxwell_WKB}

We consider the vacuum Einstein field equations with vanishing cosmological constant
\begin{equation}
    R_{\alpha \beta} - \frac{1}{2} R g_{\alpha \beta} = 0,
\end{equation}  
where $R_{\alpha \beta}$ is the Ricci tensor, $R = R\indices{^\alpha_\alpha}$ is the Ricci scalar. The Einstein field equations can be obtained as the Euler-Lagrange equations of the Einstein-Hilbert action
\begin{equation} \label{eq:EH_full}
    J({g}_{\mu \nu}) = \int_M \mathrm{d}^4 x \, \sqrt{g} \, R( {g}_{\mu \nu}).
\end{equation}
Our goal is to describe the propagation of gravitational waves, treated as a small metric perturbation around a fixed background solution of the vacuum Einstein field equations. For this purpose, in the next section we derive the linearization of the Einstein-Hilbert action and the corresponding equations for the linearized gravitational field.

Note that, we could have treated the case of a non-vanishing cosmological constant since, in the high-frequency analysis, the latter plays no role.

\subsection{Linearization of the Einstein-Hilbert action} \label{sec:linearised}

We remind here the form of the linearized Einstein-Hilbert action, see Ref. \cite{1971bicak}. For completeness, the derivation, which is often not presented in detail in the literature, is performed in Appendix \ref{app:linearization}. Let $g_{\mu \nu}$ be a solution of the Einstein field equations in vacuum
\begin{equation}
R_{\alpha\beta} = 0.
\end{equation}
We consider a Lorentzian metric $\tilde{g}_{\mu \nu}$, obtained through a small perturbation $h_{\mu \nu}$ of $g_{\mu \nu}$:
\begin{equation}
    \tilde{g}_{\mu \nu} = g_{\mu \nu} + h_{\mu \nu}.
\end{equation}
Linearizing the Einstein-Hilbert action near $g_{\mu \nu}$ as in Ref. \cite{1971bicak}, we obtain
\begin{align} 
    J(\tilde{g}_{\mu \nu}) & = J (g_{\mu \nu}) + J_{lin}(h_{\mu \nu}) + \mathcal{O}(|h|^3),
\end{align}
where
\begin{equation} \label{eq:EH_linear} 
   J_{lin}(h_{\mu \nu})= \int_M \mathrm{d}^4 x \, \sqrt{g} \, \left(- \frac{1}{2} \nabla^\gamma h^{\alpha \beta} \nabla_\gamma h_{\alpha \beta} + \frac{1}{2} \nabla_\gamma h \nabla^\gamma h - \nabla_\alpha h \nabla_\beta h^{\alpha \beta} + \nabla_\alpha h_{\gamma \beta} \nabla^\gamma h^{\alpha \beta} \right)
\end{equation}
is the action for the perturbation $h_{\mu \nu}$. Index manipulation and covariant derivatives are defined with respect to the background metric $g_{\alpha \beta}$, and $h = h_{\alpha \beta} g^{\alpha \beta}$. Integrating by parts and neglecting the boundary terms, the linearized action can be written as
\begin{equation} \label{eq:EH_linear2}
    J_{lin}(h_{\mu \nu})= \int_M \mathrm{d}^4 x \, \sqrt{g} \,  h^{\alpha \beta} \hat{D}\indices{_{\alpha \beta}^{\gamma \delta}} h_{\gamma \delta}, 
\end{equation}
where $\hat{D}\indices{_{\alpha \beta}^{\gamma \delta}}$ is the differential operator 
\begin{equation}
\hat{D}\indices{_{\alpha \beta}^{\gamma \delta}} = \frac{1}{2} \left( \delta_\alpha^\gamma \delta_\beta^\delta \nabla_\mu \nabla^\mu  - g_{\alpha \beta} g^{\gamma \delta}\nabla_\mu \nabla^\mu  + g^{\gamma \delta}\nabla_\alpha \nabla_\beta  + g_{\alpha \beta} \nabla^\gamma \nabla^\delta - \delta_\beta^\delta \nabla^\gamma\nabla_\alpha  - \delta_\alpha^\delta\nabla^\gamma \nabla_\beta  \right).
\end{equation}
The corresponding Euler-Lagrange equations are 
\begin{equation} \label{eq:MTW_Wave eqn}
    \hat{D}\indices{_{\alpha \beta}^{\gamma \delta}} h_{\gamma \delta} = 0.
\end{equation}
Introducing the trace-reverse tensor 
\begin{equation} \label{eq:trace_rev}
\breve{h} _{\alpha \beta} = h_{\alpha \beta} -\dfrac12  h g_{\alpha \beta},
\end{equation}
Eq. \eqref{eq:MTW_Wave eqn} becomes
\begin{equation}\label{eq:h_breve}
\nabla_\alpha \nabla^\alpha \breve{h}_{\mu \nu} +\nabla_\alpha \nabla_\beta  \breve{h}^{\alpha \beta} g_{\mu \nu} - \nabla^\alpha \nabla_\mu \breve{h}_{\alpha \nu} - \nabla^\alpha \nabla_\nu \breve{h}_{\alpha \mu } =0. 
\end{equation}
Taking the trace of Eq. \eqref{eq:h_breve} leads to 
\begin{equation}\label{eq:eq_trace}
\nabla_\alpha \nabla^\alpha h= 2 \nabla^\alpha \nabla^\mu \breve{h}_{\alpha \mu}.
\end{equation}

\subsection{The Lorenz gauge}\label{sec:lorenzgauge}

The Einstein field equations are gauged equations. The gauge freedom can be exploited to reduce the Einstein field equation to a hyperbolic system of equations. A detailed discussion of this reduction in the particular case of the wave gauge can be found in Refs. \cite[Section 14.1]{MR2527641} or \cite[Section 2.4]{MR1765128}. 

A similar reduction can be applied to the linearized equations \eqref{eq:MTW_Wave eqn}. The linearization of the gauge freedom of the Einstein field equations leads to the invariance of Eq. \eqref{eq:MTW_Wave eqn} by the transformation
\begin{equation}
h_{\mu\nu} \mapsto h_{\mu\nu} - \nabla_{\mu}\xi_{\nu} -  \nabla_{\nu}\xi_{\mu}, 
\end{equation}
where $\xi_\mu$ is a one-form on $M$. The gauge invariance of the linearized field equations \eqref{eq:MTW_Wave eqn} can be exploited to make these equations hyperbolic. The linearization of the wave gauge for the Einstein field equations leads to the Lorenz gauge condition for the linearized field equations \eqref{eq:MTW_Wave eqn}:
\begin{equation} \label{eq:Lorenz_gauge}
   \nabla_\alpha \breve{h}^{\alpha \beta} = \nabla_\alpha \left(  h^{\alpha \beta}  - \frac{1}{2} h g^{\alpha \beta} \right)= 0.
\end{equation}
The detailed derivation of this equation is presented in Appendix \ref{app:gauge}. Imposing the Lorenz gauge condition, Eq. \eqref{eq:MTW_Wave eqn} is reduced to the following wave equation: 
\begin{equation} \label{eq:Wave eqn with Lorenz gauge}
    \nabla^\alpha \nabla_\alpha \breve{h}_{\mu \nu} - 2R_{\nu \alpha \sigma \mu} \breve{h}^{\alpha\sigma} = 0, 
\end{equation} 
and Eq. \eqref{eq:eq_trace} for the trace of the perturbations decouples:
\begin{equation}\label{eq:eq_trace_wave}
\nabla_\alpha \nabla^\alpha h= 0.
\end{equation} 
Using the expression of $\breve{h}_{\mu \nu}$ given in Eq. \eqref{eq:trace_rev}, and using the fact that $g_{\mu \nu}$ has vanishing Ricci curvature, we obtain
\begin{equation}
    \nabla^\alpha \nabla_\alpha {h}_{\mu \nu} - 2R_{\nu \alpha \sigma \mu} {h}^{\alpha\sigma} = 0. 
\end{equation} 

\subsection{WKB Ansatz}\label{sec:wkbansatz}

We assume that the perturbation metric $h_{\alpha \beta}$ admits a WKB expansion of the form

\begin{equation} \label{eq:WKB_ansatz}
\begin{split}
    h_{\alpha \beta} (x) &= \mathrm{Re} \left[ A_{\alpha \beta}(x, k(x), \epsilon) e^{i S(x) / \epsilon} \right], \\
    A_{\alpha \beta}(x, k(x), \epsilon) &= {A_0}_{\alpha \beta}(x, k(x)) + \epsilon {A_1}_{\alpha \beta}(x, k(x)) + \mathcal{O}(\epsilon^2),
\end{split}
\end{equation}
where $S$ is a real scalar function, $A_{\alpha \beta}$ is a complex amplitude, and $\epsilon$ is a small expansion parameter. The gradient of $S$ is denoted as
\begin{equation} \label{eq:grad_S}
    k_\mu(x) = \nabla_\mu S(x).
\end{equation}
We are allowing the amplitude $A_{\alpha \beta}$ to depend on $k_\mu(x)$. This is justified by the mathematical formulation of the WKB approximation \cite{MR1806388,Emmrich1996}, where $k_\mu(x)$ determines a Lagrangian submanifold of $x \mapsto (x, k(x)) \in T^*M$, and the amplitude $A_{\alpha \beta}$ is defined on the Lagrangian submanifold.

\subsection{Assumption on the initial data}  \label{sec:initial data}

We consider a Cauchy surface in $M$, and we make the following assumptions. Firstly, the gauge condition given in Eq. \eqref{eq:Lorenz_gauge} is initially satisfied. Secondly, the trace of the perturbation $h$ vanishes initially. Equation \eqref{eq:eq_trace_wave} guarantees that this condition is conserved in the future of $\Sigma$. Finally, the gravitational waves have initially circular polarization (see Sec. \ref{sec:polarization basis}).

\section{The WKB approximation for linearized gravity} \label{sec:Higher_GO}

The WKB analysis of various field equations is generally performed by inserting the WKB ansatz directly into the field equation, followed by an analysis of the resulting terms at each order in the expansion parameter $\epsilon$. However, for the purpose of studying spin Hall effects, we find it more convenient to perform the WKB analysis by inserting the WKB ansatz into the field action. The advantages of such a variational formulation of the WKB approximation are extensively discussed in Ref. \cite{tracy2014} (see also Refs. \cite{Ruiz2015,Ruiz2017}). In particular, a similar approach proved to be effective in the derivation of the gravitational spin Hall effect of light \cite{GSHE2020}.

\subsection{Euler-Lagrange equations in the WKB approximation}\label{sec:eulerlagrangewkb}
We insert the WKB ansatz \eqref{eq:WKB_ansatz} into the linearized Einstein--Hilbert action \eqref{eq:EH_linear2}. Keeping only terms of the lowest two orders in $\epsilon$, we obtain
\begin{equation} \label{eq:eff_action}
    2\epsilon^2 J_{lin} = \int_{M}  \mathrm{d}^4 x \, \sqrt{g}  \Big[  A^{*\alpha \beta} D\indices{_{\alpha \beta}^{\gamma \delta}} A_{\gamma \delta} - \frac{i \epsilon}{2} \vnabla{}^\mu D\indices{_{\alpha \beta}^{\gamma \delta}} \left( A^{*\alpha \beta} {\nabla}_\mu A_{\gamma \delta} -  A_{\gamma \delta} {\nabla}_\mu A^{*\alpha \beta} \right) \Big] + \mathcal{O}(\epsilon^2),
\end{equation}
where
\begin{equation} \label{eq:vertical_D}
\begin{split}
    D\indices{_{\alpha \beta}^{\gamma \delta}} &= \frac{1}{2} \left( k_\mu k^\mu \delta_\alpha^\gamma \delta_\beta^\delta - k_\mu k^\mu g_{\alpha \beta} g^{\gamma \delta} + k_\alpha k_\beta g^{\gamma \delta} + k^\gamma k^\delta g_{\alpha \beta} - k_\alpha k^\gamma \delta_\beta^\delta - k_\beta k^\gamma \delta_\alpha^\delta \right), \\
    \vnabla{}^\mu D\indices{_{\alpha \beta}^{\gamma \delta}} &= k^\mu \delta_\alpha^\gamma \delta_\beta^\delta - k^\mu g_{\alpha \beta} g^{\gamma \delta} + k_{(\alpha} \delta_{\beta)}^\mu g^{\gamma \delta} + k^{(\gamma} g^{\delta) \mu} g_{\alpha \beta} - k_{(\alpha} \delta_{\beta)}^\delta g^{\gamma \mu} - k^\gamma \delta_{(\alpha}^\mu \delta_{\beta)}^\delta, \\
    \vnabla{}^\mu \vnabla{}^\nu D\indices{_{\alpha \beta}^{\gamma \delta}} &= g^{\mu \nu} \delta_\alpha^\gamma \delta_\beta^\delta - g^{\mu \nu} g_{\alpha \beta} g^{\gamma \delta} + \delta^\mu_{(\alpha} \delta_{\beta)}^\nu g^{\gamma \delta} + g^{\mu (\gamma} g^{\delta) \nu} g_{\alpha \beta} - \delta^\mu_{(\alpha} \delta_{\beta)}^\delta g^{\gamma \nu} - g^{\gamma \mu} \delta_{(\alpha}^\nu \delta_{\beta)}^\delta
\end{split}  
\end{equation}
In the above equation, $D\indices{_{\alpha \beta}^{\gamma \delta}}$ represents the symbol of the operator $ \hat{D}\indices{_{\alpha \beta}^{\gamma \delta}}$, and $\vnabla{}^\mu =\frac{\partial}{\partial k_\mu}$ denotes the vertical derivative (see Ref. \cite[Appendix A]{GSHE2020} for the definition of horizontal and vertical derivatives). Formally, up to the expression of the symbol $D\indices{_{\alpha \beta}^{\gamma \delta}}$, the effective action \eqref{eq:eff_action} is of the same form as the effective action obtained in the electromagnetic case \cite[Eq. (3.3)]{GSHE2020}.

The effective action \eqref{eq:eff_action} depends on $S(x)$, $A_{\alpha \beta}(x,\nabla S)$ and ${A^*}^{\alpha \beta}(x,\nabla S)$, and the variation can be performed as in \cite[Appendix B]{GSHE2020}. The resulting Euler--Lagrange equations are
\begin{align}
    \label{eq:first EOM} D\indices{_{\alpha \beta}^{\gamma \delta}} A_{\gamma \delta} - i \epsilon (\vnabla{}^\mu D\indices{_{\alpha \beta}^{\gamma \delta}}) \nabla_\mu A_{\gamma \delta} - \frac{i \epsilon}{2} (\nabla_\mu \vnabla{}^\mu D\indices{_{\alpha \beta}^{\gamma \delta}}) A_{\gamma \delta} = \mathcal{O}(\epsilon^2) \\
     \label{eq:second EOM} D\indices{_{\alpha \beta}^{\gamma \delta}} {A^*}^{\alpha \beta} + i \epsilon (\vnabla{}^\mu D\indices{_{\alpha \beta}^{\gamma \delta}}) \nabla_\mu {A^*}^{\alpha \beta} + \frac{i \epsilon}{2} (\nabla_\mu \vnabla{}^\mu D\indices{_{\alpha \beta}^{\gamma \delta}}) {A^*}^{\alpha \beta} = \mathcal{O}(\epsilon^2) \\
     \label{eq:third EOM} \nabla_\mu \bigg[ (\vnabla{}^\mu D\indices{_{\alpha \beta}^{\gamma \delta}}) {A^*}^{\alpha \beta} A_{\gamma \delta} -\frac{i \epsilon}{2} (\vnabla{}^\mu \vnabla{}^\nu D\indices{_{\alpha \beta}^{\gamma \delta}}) \left( A^{*\alpha \beta} {\nabla}_\nu A_{\gamma \delta} -  A_{\gamma \delta} {\nabla}_\nu A^{*\alpha \beta} \right) \bigg] = \mathcal{O}(\epsilon^2) 
\end{align}
In the above equations, the symbol $D\indices{_{\alpha \beta}^{\gamma \delta}}$ and its vertical derivatives are evaluated at the phase space point $(x, p) = (x, k)$.

\subsection{WKB approximation of the Lorenz gauge}\label{sec:consequencelorenzgauge}

In order to remove unwanted pure gauge degrees of freedom, the Euler-Lagrange equations \eqref{eq:first EOM}-\eqref{eq:third EOM} should be supplemented with additional equations. For this purpose, we impose the Lorenz gauge condition  on the metric perturbation $h_{\alpha \beta}$. The WKB approximation of the Lorenz gauge condition is obtained by inserting the WKB ansatz \eqref{eq:WKB_ansatz} into Eq. \eqref{eq:Lorenz_gauge}. At the lowest order in $\epsilon$, we obtain
\begin{equation} \label{eq:Lorenz_0}
      k^\alpha {A_0}_{\alpha \mu} = \frac{1}{2} k_\mu A_0,
\end{equation}
and at $\mathcal{O}(\epsilon^0)$ we obtain
\begin{align} \label{eq:Lorenz_1}
    \nabla^\alpha {A_0}_{\alpha \mu} + i k^\alpha {A_1}_{\alpha \mu} = \frac{1}{2} ( \nabla_\mu A_0 + i k_\mu A_1),
\end{align}
where $A_0 = g^{\alpha \beta} {A_0}_{\alpha \beta}$ and $A_1 = g^{\alpha \beta} {A_1}_{\alpha \beta}$. These equations can also be supplemented by the corresponding complex conjugate equations.

\subsection{Equations at order $\epsilon^0$}\label{sec:0thGO}

Keeping only terms of order $\epsilon^0$, Equations \eqref{eq:first EOM}-\eqref{eq:third EOM} reduce to
\begin{align}
     \label{0th order first EOM} D\indices{_{\alpha \beta}^{\gamma \delta}} {A_0}_{\gamma \delta} = 0 \\
     \label{0th order second EOM}D\indices{_{\alpha \beta}^{\gamma \delta}} {{A_0}^*}^{\alpha \beta} = 0 \\
     \label{0th order third EOM} \nabla_\mu \bigg[ (\vnabla{}^\mu D\indices{_{\alpha \beta}^{\gamma \delta}}) {{A_0}^*}^{\alpha \beta} {A_0}_{\gamma \delta} \bigg] = 0
\end{align}
Since Equations \eqref{0th order first EOM} and \eqref{0th order second EOM} are related by complex conjugation, it is enough to analyze only one of them. Using the definition of the symbol $D\indices{_{\alpha \beta}^{\gamma \delta}}$, Equation \eqref{0th order first EOM} can be written as 
\begin{equation} \label{eq:First EOM expanded}
    \frac{1}{2} \left( k_\mu k^\mu \delta_\alpha^\gamma \delta_\beta^\delta - k_\mu k^\mu g_{\alpha \beta} g^{\gamma \delta} + k_\alpha k_\beta g^{\gamma \delta} + k^\gamma k^\delta g_{\alpha \beta} - k_\alpha k^\gamma \delta_\beta^\delta - k_\beta k^\gamma \delta_\alpha^\delta \right) {A_0}_{\gamma \delta} = 0
\end{equation}
This equation admits nontrivial solutions if and only if ${A_0}_{\gamma \delta}$ is in the kernel of the tensor $D\indices{_{\alpha \beta}^{\gamma \delta}}$. The kernel of $D\indices{_{\alpha \beta}^{\gamma \delta}}$ is discussed in detail in Appendix \ref{app:symbol}. By imposing the Lorenz gauge condition \eqref{eq:Lorenz_0} in Eq. \eqref{eq:First EOM expanded}, we obtain
\begin{equation}
    k_\mu k^\mu \bigg( {A_0}_{\alpha \beta} -\frac{1}{2} g_{\alpha \beta} A_0 \bigg) = 0
\end{equation}
This equation can only be satisfied if either $k_\mu k^\mu = 0$ or ${A_0}_{\alpha \beta} -\frac{1}{2} g_{\alpha \beta} A_0 = 0$. However, taking ${A_0}_{\alpha \beta} -\frac{1}{2} g_{\alpha \beta} A_0 = 0$ implies that ${A_0}_{\alpha \beta} = 0$. Discarding this trivial solution, we are left with the dispersion relation 
\begin{equation} \label{eq:disp0}
    k_\mu k^\mu = 0,
\end{equation}
which is a well-known result of geometrical optics. Furthermore, since $k_\mu$ is the gradient of a scalar function, it satisfies
\begin{equation} \label{eq:sym_nabla_k}
    \nabla_\mu k_\alpha = \nabla_\alpha k_\mu.
\end{equation}
Using this property, together with the dispersion relation \eqref{eq:disp0}, we can derive the geodesic equation for $k_\mu$:
\begin{equation} \label{eq:geodesic}
    k^\nu \nabla_\nu k_\mu = 0.
\end{equation}
Imposing the Lorenz gauge condition \eqref{eq:Lorenz_0} in Equation \eqref{0th order third EOM}, we obtain
\begin{equation} \label{eq:transp0}
    \nabla_\mu \bigg[ k^\mu \left( {{A_0}^*}^{\alpha \beta} {A_0}_{\alpha \beta} - \frac{1}{2} {A_0}^* A_0 \right) \bigg] = 0. 
\end{equation}
This equation represents a transport equation for the intensity $\mathcal{I}_0 = {{A_0}^*}^{\alpha \beta} {A_0}_{\alpha \beta} - \frac{1}{2} {A_0}^* A_0$, which is another well-known result of geometrical optics.

\subsection{Equations at order $\epsilon^1$} \label{sec:GO1storder}

We continue the WKB analysis by taking equations \eqref{0th order first EOM} and \eqref{0th order second EOM} at order $\epsilon^1$ only:
\begin{align} \label{eq:1st order first EOM}
    D\indices{_{\alpha \beta}^{\gamma \delta}} {A_1}_{\gamma \delta} - i (\vnabla{}^\mu D\indices{_{\alpha \beta}^{\gamma \delta}}) \nabla_\mu {A_0}_{\gamma \delta} - \frac{i}{2} (\nabla_\mu \vnabla{}^\mu D\indices{_{\alpha \beta}^{\gamma \delta}}) {A_0}_{\gamma \delta} &= 0, \\
    D\indices{_{\alpha \beta}^{\gamma \delta}} {{A_1}^*}^{\alpha \beta} + i (\vnabla{}^\mu D\indices{_{\alpha \beta}^{\gamma \delta}}) \nabla_\mu {{A_0}^*}^{\alpha \beta} + \frac{i}{2} (\nabla_\mu \vnabla{}^\mu D\indices{_{\alpha \beta}^{\gamma \delta}}) {{A_0}^*}^{\alpha \beta} &= 0.  \label{eq:1st order second EOM}
\end{align}
We can simplify these equations by imposing the Lorenz gauge condition \eqref{eq:Lorenz_0} and \eqref{eq:Lorenz_1}, and by using equations \eqref{eq:disp0} and \eqref{eq:sym_nabla_k}. We obtain
\begin{align}
    \label{eq:first order first EOM simplified} k^\mu \nabla_\mu \bigg( {A_0}_{\alpha \beta} - \frac{1}{2} g_{\alpha \beta} A_0 \bigg) + \frac{1}{2} \bigg( {A_0}_{\alpha \beta} - \frac{1}{2} g_{\alpha \beta} A_0 \bigg) \nabla_\mu k^\mu &= 0, \\
      \label{eq:first order second EOM simplified}k^\mu \nabla_\mu \bigg( {{A_0}^*}^{\alpha \beta} - \frac{1}{2} g^{\alpha \beta} {A_0}^* \bigg) + \frac{1}{2} \bigg( {{A_0}^*}^{\alpha \beta} - \frac{1}{2} g^{\alpha \beta} {A_0}^* \bigg) \nabla_\mu k^\mu &= 0.
\end{align}
Furthermore, using the lowest-order intensity $\mathcal{I}_0$, we can write the amplitude tensors in the following way:
\begin{equation} \label{eq:a0_def}
    {A_0}_{\alpha \beta} - \frac{1}{2} g_{\alpha \beta} A_0 = \sqrt{\mathcal{I}_0} {a_0}_{\alpha \beta}, \qquad    
    {{A_0}^*}^{\alpha \beta} - \frac{1}{2} g^{\alpha \beta} {A_0}^* = \sqrt{\mathcal{I}_0} {{a_0}^*}^{\alpha \beta},
\end{equation}
where ${a_0}_{\alpha \beta}$ is a complex tensor, describing the polarization of the gravitational wave. Note that, due to the Lorenz gauge condition \eqref{eq:Lorenz_0}, the polarization tensor ${a_0}_{\alpha \beta}$ satisfies the orthogonality condition
\begin{equation} \label{eq:orthogonality of k and polarization tensor}
    k^\alpha {a_0}_{\alpha \beta} = 0.
\end{equation}
Using the transport equation \eqref{eq:transp0}, Eqs. \eqref{eq:first order first EOM simplified} and \eqref{eq:first order second EOM simplified} reduce to
\begin{equation}\label{eq:polarization}
    k^\mu \nabla_\mu {a_0}_{\alpha \beta} = k^\mu \nabla_\mu {{a_0}^*}^{\alpha \beta} = 0.
\end{equation}
The parallel propagation of the complex polarization tensor ${a_0}_{\alpha \beta}$ along $k^\mu$ is another well-known result of the geometrical optics approximation.

\subsection{The polarization tensor in a null tetrad} \label{sec:polarization basis}

The properties of the polarization tensor ${a_0}_{\alpha \beta}$ become more transparent when expressed in terms of a null tetrad adapted to $k_\alpha$. Working with the metric signature $(-,+,+,+)$, we establish a set of four complex null vectors $\{k_\alpha,n_\alpha,m_\alpha,\bar{m}_\alpha\}$ at each point in space-time , which satisfy the following orthogonality relations: 
\begin{equation}\label{eq:NPtetrad}
\begin{split}
    m_\alpha \bar{m}^\alpha = 1, \qquad k_\alpha n^\alpha = -1, \\
    k_\alpha k^\alpha = n_\alpha n^\alpha = m_\alpha m^\alpha = \bar{m}_\alpha \bar{m}^\alpha &= 0, \\
    k_\alpha m^\alpha = k_\alpha \bar{m}^\alpha = n_\alpha m^\alpha = n_\alpha \bar{m}^\alpha &= 0.
\end{split}
\end{equation}
Since the polarization tensor ${a_0}_{\mu \nu}$ is symmetric, it can have at most ten independent components. However, due to the orthogonality condition \eqref{eq:orthogonality of k and polarization tensor}, we are left with only six independent components. Using the null tetrad, we can write the polarization tensor as 
\begin{equation} \label{eq:form of the polarization tensor}
\begin{split} 
    {a_0}_{\mu \nu}= z_1 m_\mu m_\nu + z_2 \bar{m}_\mu \bar{m}_\nu +  z_3 m_{(\mu} \bar{m}_{\nu)} + z_4  k_\mu k_\nu + z_5 k_{(\mu} m_{\nu)} + z_6 k_{(\mu} \bar{m}_{\nu)},
\end{split}
\end{equation} 
where $z_i$ are complex scalar functions. Inserting this expansion of the polarization tensor into the parallel transport equation \eqref{eq:polarization}, and making use of the orthogonality relations \eqref{eq:NPtetrad}, we obtain the following transport equations for the scalar functions $z_i$:
\begin{subequations}  \label{eq:z_transport}
\begin{align}
    k^\alpha \nabla_\alpha z_1 &= -2 z_1 \bar{m}^\mu k^\alpha \nabla_\alpha m_\mu, \\
    k^\alpha \nabla_\alpha z_2 &= -2 z_2 m^\mu k^\alpha \nabla_\alpha \bar{m}_\mu, \\
    k^\alpha \nabla_\alpha z_3 &= 0, \\
    k^\alpha \nabla_\alpha z_4 &= -(z_5 m^\mu+z_6\bar{m}^\mu) k^\alpha \nabla_\alpha n_\mu, \\
    k^\alpha \nabla_\alpha z_5 &= -(z_3 \bar{m}^\mu + 2 z_1 m^\mu) k^\alpha\nabla_\alpha n_\mu - z_5 \bar{m}^\nu k^\alpha\nabla_\alpha m_\nu, \\
    k^\alpha \nabla_\alpha z_6 &= -(z_3 m^\mu + 2 z_2  \bar{m}^\mu) k^\alpha\nabla_\alpha n_\mu - z_6 m^\nu k^\alpha \nabla_\alpha \bar{m}_\nu.
\end{align}
\end{subequations}
The transport equations for $z_1$, $z_2$ and $z_3$ are decoupled. Furthermore, the evolution of the trace of ${a_0}_{\mu \nu}$ is described by $z_3$, which is covariantly constant along $k^\alpha$, and its value will be fixed by the choice of initial conditions. As mentioned in Sec. \ref{sec:initial data}, we consider initial data such that the metric perturbation is initially traceless. Thus, we impose $z_3 = 0$. The other components, $z_4$, $z_5$ and $z_6$, describe the evolution of pure gauge degrees of freedom, which were not fixed by imposing the Lorenz gauge. It is shown in Appendix \ref{app:symbol} that the components of the metric perturbation proportional to $z_4$, $z_5$ and $z_6$ do not contribute, at the lowest order in $\epsilon$, to the Riemann tensor. They are in that sense pure gauge. 

The non pure-gauge degrees of freedom, describing the polarization of the metric perturbation are represented by the terms proportional to the complex scalar functions $z_1$ and $z_2$. The tensors $m_\mu m_\nu$ and $\bar{m}_\mu \bar{m}_\nu$ represent a circular polarization basis for linearized metric perturbations, analogue to the circular polarization basis covectors $m_\mu$ and $\bar{m}_\mu$ used in the description of electromagnetic waves (a detailed comparison between the polarization of electromagnetic and gravitational waves can be found in \cite[Sec. 35.6]{MTW}). By picking initial data such that the metric perturbation is initially traceless (which is equivalent to $z_3 = 0$), Eq. \eqref{eq:a0_def} implies that
\begin{equation}
    {a_0}^{* \mu \nu} {a_0}_{\mu \nu} = z^*_1 z_1 + z^*_2 z_2 = 1.
\end{equation}
This relation restricts $(z_1, z_2) \in \mathbb{C}^2$ to the unit $3$-sphere $S^3$. Furthermore, $(z_1, z_2)$ and $(e^{i \phi} z_1, e^{i \phi} z_2)$ (for any real $\phi$), represent the same polarization state. Thus, the space of physically distinguishable polarization states is the complex projective line $\mathds{C}\mathrm{P}^1 = S^3/U(1) = S^2$ (in optics, this is called the Poincare sphere; see Refs \cite[Sec. 1.4.2]{born1980} \cite[Sec 5.2]{bengtsson2017}). 

The transport equations for $z_1$ and $z_2$ have the same form as in the electromagnetic case \cite[Eq. (3.36)]{GSHE2020}, the only difference being a factor of $2$, which corresponds to the fact that here we are dealing with a spin-$2$ field, instead of the electromagnetic field, which is a spin-$1$ field. As in Ref. \cite{GSHE2020}, it is convenient to rewrite the transport equations for $z_1$ and $z_2$ in terms of the Berry connection. First, we should remember that the covectors $m_\alpha$ and $\bar{m}_\alpha$ are functions of $x$ and $k(x)$, because of the orthogonality relations given in Eq. \eqref{eq:NPtetrad}. Thus, we must carefully apply the chain rule when taking covariant derivatives of $m_\alpha$ and $\bar{m}_\alpha$:
\begin{equation}
\begin{split}
    k^\mu \nabla_\mu m_\alpha &= k^\mu \nabla_\mu \left[ m_\alpha (x, k) \right] \\
    &= k^\mu \left( \hnabla{}_\mu m_\alpha \right) (x, k) + k^\mu \left( \nabla_\mu k_\nu \right) \left(\vnabla{}^\nu m_\alpha \right) (x, k) \\
    &= k^\mu \hnabla{}_\mu m_\alpha,
\end{split}
\end{equation}
where $\hnabla_\mu$ is the horizontal derivative, defined in Ref. \cite[Appendix A]{GSHE2020}. As in the electromagnetic case, the scalar functions $z_1$ and $z_2$ can be encoded in a $2$-dimensional unit complex vector, which is analogous to the Jones vector used in optics \cite{optics_book,Bliokh2009,Ruiz2015,Ruiz2017}:
\begin{equation}
    z = \begin{pmatrix} z_1 \\ z_2 \end{pmatrix}, \qquad z^\dagger =  \begin{pmatrix} z_1^* & z_2^* \end{pmatrix}.
\end{equation}
The transport equations for $z_1$ and $z_2$ can be rewritten as
\begin{equation}
    k^\mu \nabla_\mu z = 2 i k^\mu B_\mu \sigma_3 z,
\end{equation}
where $\sigma_3$ is the third Pauli matrix, 
\begin{equation}
    \sigma_3 = \begin{pmatrix} 1 & 0 \\ 0 & -1 \end{pmatrix},
\end{equation}
and $B_\mu$ is the Berry connection
\begin{equation} \label{eq:B_connection}
    B_\mu(x, k) = \frac{i}{2} \left( \bar{m}^{\alpha} \hnabla_\mu {m}_\alpha - {m}_\alpha \hnabla_\mu \bar{m}^{\alpha} \right) = i \bar{m}^{\alpha} \hnabla_\mu {m}_\alpha.
\end{equation}
The Berry connection has the same definition as in the electromagnetic case \cite{GSHE2020}. The Berry phase can be defined by considering a worldline $x^\mu(\tau)$, with $\dot{x}^\mu = k^\mu$. Then, by restricting $z$ to the worldline $x^\mu(\tau)$, we obtain
\begin{equation} \label{eq:z_dot}
    \dot z = 2 i k^\mu B_\mu \sigma_3 z.
\end{equation}
This equation can be integrated along the worldline $x^\mu(\tau)$ as
\begin{equation} \label{eq:z_integration}
    z(\tau) = \begin{pmatrix} e^{ 2 i \gamma(\tau)} & 0 \\ 0 & e^{- 2 i \gamma(\tau)} \end{pmatrix} z(0),
\end{equation}
and we obtain the Berry phase $\gamma$ as
\begin{equation}
    \gamma(\tau_1) = \int_{\tau_0}^{\tau_1} d \tau k^\mu B_\mu.
\end{equation}
Using equation \eqref{eq:z_dot}, we can show that the following quantities are conserved along $k^\mu$: 
\begin{equation}
\begin{split}
1 &= z_1^* z_1 + z_2^* z_2 = z^\dag z , \\
s &= 2 ( z_1^* z_1 - z_2^* z_2 ) = 2 z^\dagger \sigma_3 z.
\end{split}
\end{equation} 
Based on our assumptions on the initial conditions, given in Sec. \ref{sec:initial data}, we only consider metric perturbations which are initially circularly polarized. This corresponds to
\begin{equation}
    z(0) =  \begin{pmatrix} 1 \\ 0 \end{pmatrix}   \qquad \text{or} \qquad z(0) =  \begin{pmatrix} 0 \\ 1 \end{pmatrix}.
\end{equation}
Thus, we have $s=\pm 2$, depending on the choice of the initial polarization state. Here, the parameter $s$ represents the helicity of the metric perturbation.

\subsection{Effective dispersion relation} \label{sec:effectivedispersion}

The results derived so far are based on a standard approach to the WKB analysis, by imposing that terms at different orders in $\epsilon$ in the Euler-Lagrange equations \eqref{eq:first EOM}-\eqref{eq:third EOM} vanish separately. With this approach, we derived the well-known geometrical optics results: the dispersion relation \eqref{eq:disp0} and the transport equation for the polarization tensor \eqref{eq:polarization}. While the dynamics of the polarization tensor in Eq. \eqref{eq:polarization} depends on $k_\mu$, and, hence, on the dispersion relation \eqref{eq:disp0}, there is no backreaction from the dynamics of the polarization tensor onto the dispersion relation \eqref{eq:polarization} and onto $k_\mu$. In other words, the standard geometrical optics approach does not take into account all the possible spin-orbit interactions between the external and internal degrees of freedom, here represented by the wave vector $k_\mu$ and polarization tensor ${a_0}_{\mu \nu}$. 

In the derivation of the spin Hall effect, as observed in Ref. \cite{GSHE2020} (see also Ref. \cite{SOI_review}), it is essential to gather terms related to geometrical optics and terms involving the polarization. This is the so-called spin-orbit coupling. This can be achieved by collating the separately satisfied Eqs. \eqref{eq:first EOM}--\eqref{eq:third EOM} into one quantity depending on powers of $\epsilon$ at order $0$ and $1$, and vanishing at order $\mathcal{O}(\epsilon^2)$.

Starting with Eqs. \eqref{eq:first EOM}--\eqref{eq:third EOM}, an effective dispersion relation can be derived in the in the following way. We contract Eq. \eqref{eq:first EOM} with ${A^*}^{\alpha \beta}$ and Eq. \eqref{eq:second EOM} with $A_{\gamma \delta}$. Adding these equations together, we obtain
\begin{equation} \label{eq:effective_disp_1}
    D\indices{_{\alpha \beta}^{\gamma \delta}} {A^*}^{\alpha \beta} {A}_{\gamma \delta} - \frac{i \epsilon}{2} \left( \vnabla{}^\mu D\indices{_{\alpha \beta}^{\gamma \delta}} \right) \left( {A^*}^{\alpha \beta} \nabla_\mu {A}_{\gamma \delta} - {A}_{\gamma \delta} \nabla_\mu {A^*}^{\alpha \beta} \right) = \mathcal{O}(\epsilon^2).
\end{equation}
Using $A_{\alpha \beta} = {A_0}_{\alpha \beta} + \epsilon {A_1}_{\alpha \beta} + \mathcal{O}(\epsilon^2)$, the Lorenz gauge condition given in Eqs. \eqref{eq:Lorenz_0} and \eqref{eq:Lorenz_1}, as well as Eq. \eqref{eq:vertical_D}, we can rewrite the above equation as
\begin{equation} \label{eq:Dispersion relation 3}
\begin{split}
    \frac{1}{2}& k^\mu k_\mu \bigg[ \mathcal{I}_0 + \epsilon \left( {A_0}_{\alpha \beta} {{A_1}^*}^{\alpha \beta} + {{A_0}^*}^{\alpha \beta} {A_1}_{\alpha \beta} - \frac{1}{2} A_0 {A_1}^* - \frac{1}{2} A_1 {A_0}^* \right) \bigg] \\
    &-\frac{i \epsilon}{2} k^\mu \bigg[ {{A_0}^*}^{\gamma \delta} \nabla_\mu {A_0}_{\gamma \delta} - {A_0}_{\gamma \delta} \nabla_\mu {{A_0}^*}^{\gamma \delta} + \frac{1}{2} \left( {A_0}^* \nabla_\mu A_0 - A_0 \nabla_\mu {A_0}^* \right) \bigg] = \mathcal{O}(\epsilon)^2 
\end{split}
\end{equation}
The above equation can be further simplified by introducing the $\mathcal{O}(\epsilon)^1$ intensity as
\begin{equation} \label{eq:first order intensity}
\begin{split}
      \mathcal{I}_1 &= \bigg( {A}_{\alpha \beta} {{A}^*}^{\alpha \beta} - \frac{1}{2} A {A}^* \bigg) + \mathcal{O}(\epsilon)^2\\
      &= \mathcal{I}_0 + \epsilon \left( {A_0}_{\alpha \beta} {{A_1}^*}^{\alpha \beta} + {{A_0}^*}^{\alpha \beta} {A_1}_{\alpha \beta} - \frac{1}{2} A_0 {A_1}^* - \frac{1}{2} A_1 {A_0}^* \right) + \mathcal{O}(\epsilon)^2.
\end{split}
\end{equation}
Then, we can rewrite the amplitude as 
\begin{equation}
    A_{\alpha \beta} = \sqrt{\mathcal{I}_1} a_{\alpha \beta} = \sqrt{\mathcal{I}_1} ( {a_0}_{\alpha \beta} + \epsilon {a_1}_{\alpha \beta} ) + \mathcal{O}(\epsilon)^2, 
\end{equation}
and from Eq. \eqref{eq:Dispersion relation 3} we obtain
\begin{equation} \label{eq:eff_disp}
    \frac{1}{2}k_\mu k^\mu - \frac{i\epsilon}{2} k^\mu \left( {{a_0}^*}^{\alpha \beta} \nabla_\mu {a_0}_{\alpha \beta} - {a_0}_{\alpha \beta} \nabla_\mu {{a_0}^*}^{\alpha \beta} \right) = \mathcal{O}(\epsilon^2).
\end{equation}
This represents an effective dispersion relation, containing $\mathcal{O}(\epsilon)$ corrections to the geometrical optics equation \eqref{eq:disp0}. We can also introduce the notation
\begin{equation}
    K_\mu = k_\mu - \frac{i\epsilon}{2} \left( {{a_0}^*}^{\alpha \beta} \nabla_\mu {a_0}_{\alpha \beta} - {a_0}_{\alpha \beta} \nabla_\mu {{a_0}^*}^{\alpha \beta} \right)
\end{equation}
and rewrite the effective dispersion relation as
\begin{equation}
    \frac{1}{2} K_\mu K^\mu  = \mathcal{O}(\epsilon^2).
\end{equation}
In a similar way, starting with \eqref{eq:third EOM}, and considering $A_{\alpha \beta} = {A_0}_{\alpha \beta} + \epsilon {A_1}_{\alpha \beta} + \mathcal{O}(\epsilon^2)$, the Lorenz gauge condition given in Eqs. \eqref{eq:Lorenz_0} and \eqref{eq:Lorenz_1}, as well as Eq. \eqref{eq:vertical_D}, we obtain
\begin{equation} \label{eq:eff_transp}
    \nabla_\mu \Bigg\{ \mathcal{I}_1 \left[ k^\mu - \frac{i\epsilon}{2} g^{\mu \nu} \left( {{a_0}^*}^{\alpha \beta} \nabla_\nu {a_0}_{\alpha \beta} - {a_0}_{\alpha \beta} \nabla_\nu  {{a_0}^*}^{\alpha \beta} \right) \right] \Bigg\} = \nabla_\mu \left( \mathcal{I}_1 K^\mu \right) = \mathcal{O}(\epsilon^2).
\end{equation}
This is an effective transport equation for the intensity $\mathcal{I}_1$, which includes $\mathcal{O}(\epsilon)$ corrections to the geometrical optics equation \eqref{eq:transp0}.

\section{Effective ray equations}\label{sec:effectiveray} 

The transition from the WKB approximation of a field theory to an effective point-particle description can be realized by treating the dispersion relation as a Hamilton-Jacobi equation for the phase function \cite[Sec. 46]{Arnold_book}. It has also been argued in Refs. \cite[Box 25.3]{MTW} \cite[Sec. II]{Gerlach1969} that the physical interpretation of the effective point-particle description provided by solving the Hamilton-Jacobi equation is related to the principle of constructive interference. One can define a localized wave packet by considering a superposition of WKB wave functions with slightly different wave vectors. The peak of intensity of this superposition occurs where the waves interfere constructively and coincides with the ray trajectories given by the effective point-particle description.

At the lowest order in $\epsilon$, we obtained in Eq. \eqref{eq:disp0} the dispersion relation
\begin{equation}
    \frac{1}{2} g^{\mu \nu} k_\mu k_\nu = 0,
\end{equation}
where $k_\mu = \nabla_\mu S$. This can be viewed as a Hamilton-Jacobi equation, which is a nonlinear first-order partial differential equation for the phase function $S$. We can solve the Hamilton-Jacobi equation by using the method of characteristics \cite[Sec. 46]{Arnold_book}. This is done by defining a Hamiltonian function $H(x, p)$ on $T^*M$, related to the dispersion relation by
\begin{equation} \label{eq:HJ_0}
    H \left(x, \nabla S \right) = \frac{1}{2}g^{\mu \nu} k_\mu k_\nu = 0.
\end{equation}
In this case, the Hamiltonian function is
\begin{equation} \label{eq:H_0}
    H(x, p) = \frac{1}{2}g^{\mu \nu} p_\mu p_\nu,
\end{equation}
where $p_\mu$ is a general covector on $T^*M$, unlike $k_\mu$, which is a gradient of a scalar function. The effective point-particle description is given by Hamilton's equations
\begin{align} 
    \dot{x}^\mu &= \frac{\partial H}{\partial p_\mu} = g^{\mu \nu} p_\nu, \label{eq:EOM_0_x}\\
    \dot{p}_\mu &= -\frac{\partial H}{\partial x^\mu} = -\frac{1}{2} \partial_\mu g^{\alpha \beta} p_\alpha p_\beta. \label{eq:EOM_0_p}
\end{align}
Given a set of ray trajectories $\{x^\mu(\tau), p_\mu(\tau)\}$ representing a solution of Hamilton's equations, we can obtain a solution of the Hamilton--Jacobi equation as \cite[Sec. 46]{Arnold_book}
\begin{equation}
    S(x^\mu(\tau_1), p_\mu(\tau_1)) = \int_{\tau_0}^{\tau_1} d \tau \left[ \dot{x}^\mu p_\mu - H(x, p) \right] + \text{const}. 
\end{equation}
Thus, at the lowest order in $\epsilon$ of the WKB approximation, we have obtained an effective point-particle description in terms of Hamilton's equations \eqref{eq:EOM_0_x} and \eqref{eq:EOM_0_p}. These are the geodesic equations of the underlying spacetime.

In order to describe spin Hall effects, higher-order terms in the WKB analysis must be taken into account. This can be achieved by considering the effective dispersion relation obtained in Eq. \eqref{eq:eff_disp}:
\begin{equation}
    \frac{1}{2}k_\mu k^\mu - \frac{i\epsilon}{2} k^\mu \left( {{a_0}^*}^{\alpha \beta} \nabla_\mu {a_0}_{\alpha \beta} - {a_0}_{\alpha \beta} \nabla_\mu {{a_0}^*}^{\alpha \beta} \right) = \mathcal{O}(\epsilon^2).
\end{equation}
Our aim is to treat this relation as an effective Hamilton-Jacobi equation, and to explore the corresponding effective point-particle description. Using the expansion of the polarization tensor ${a_0}_{\alpha \beta}$, given in Eq. \eqref{eq:form of the polarization tensor}, we can rewrite the effective dispersion relation as 
\begin{equation} \label{eq:eff_disp_2}
    \frac{1}{2} g^{\mu \nu} k_\mu k_\nu - \frac{i\epsilon}{2} k^\mu \left( z^\dagger \partial_\mu z - \partial_\mu z^\dagger z \right) 
    - \epsilon s k^\mu B_\mu = \mathcal{O}(\epsilon^2),
\end{equation}
where $B_\mu = B_\mu(x, k)$ is the Berry connection defined in Eq. \eqref{eq:B_connection}, and $s = \pm 2$, depending on the initial state of circular polarization. Note that, except for the different value of the constant $s$, we have obtained the same effective dispersion relation as in the electromagnetic case \cite[Eq. (4.12)]{GSHE2020}. Using Eq. \eqref{eq:z_integration}, we can rewrite the second term in Eq. \eqref{eq:eff_disp_2} in terms of the Berry phase $\gamma$:
\begin{equation}
    -\frac{i\epsilon}{2} k^\mu \left( z^\dagger \partial_\mu z - \partial_\mu z^\dagger z \right) = \epsilon s k^\mu \partial_\mu \gamma.
\end{equation}
Using the Berry phase, we can define an effective phase function $\tilde{S} = S + \epsilon s \gamma$ and an effective wave vector $\nabla_\mu \tilde{S} = \tilde{k}_\mu = k_\mu + \epsilon s \nabla_\mu \gamma$. Then, the effective dispersion relation can be written as
\begin{equation}
    \frac{1}{2} g^{\mu \nu} \tilde{k}_\mu \tilde{k}_\nu - \epsilon s \tilde{k}^\mu B_\mu = \mathcal{O}(\epsilon^2),
\end{equation}
This equation can be considered as an effective Hamilton-Jacobi equation for the effective phase function $\tilde{S}$. Since circularly polarized WKB metric perturbations are of the form 
\begin{equation}
    h_{\alpha \beta} = \mathrm{Re} \left[ \sqrt{\mathcal{I}} m_\alpha m_\beta e^{i \gamma} e^{i S(x) / \epsilon} \right] \qquad \text{or} \qquad h_{\alpha \beta} =\text{Re}  \left[ \sqrt{\mathcal{I}} \bar{m}_\alpha \bar{m}_\beta e^{-i \gamma} e^{i S(x) / \epsilon} \right],
\end{equation}
the effective phase function $\tilde{S}$ represents the overall phase factor of the WKB ansatz, up to order $\mathcal{O}(\epsilon^2)$. As in the previous case, we solve the effective Hamilton-Jacobi equation for the unknown $\tilde{S}$ by using the method of characteristics. We are seeking a Hamiltonian function $H(x, p)$ on $T^*M$, related to the effective dispersion relation by 
\begin{equation}
    H\left(x, \nabla \tilde{S} \right) = \frac{1}{2} g^{\mu \nu} \tilde{k}_\mu \tilde{k}_\nu - \epsilon s \tilde{k}^\mu B_\mu = \mathcal{O}(\epsilon^2).
\end{equation}
In this case, the Hamiltonian function is
\begin{equation} \label{eq:H_canonical}
    H(x, p) = \frac{1}{2}g^{\mu \nu} p_\mu p_\nu -\epsilon s g^{\mu \nu} p_\mu B_\nu(x, p),
\end{equation}
and the effective point-particle description is given by Hamilton's equations
\begin{align}
    \dot{x}^\mu &= \frac{\partial H}{\partial p_\mu} = g^{\mu \nu} p_\nu - \epsilon s \left( B^\mu + p^\alpha \vnabla{}^\mu B_\alpha \right), \label{eq:EOM_1_x}\\
    \dot{p}_\mu &= -\frac{\partial H}{\partial x^\mu} = -\frac{1}{2} \partial_\mu g^{\alpha \beta} p_\alpha p_\beta + \epsilon s p_\alpha \left( \partial_\mu g^{\alpha \beta} B_\beta + g^{\alpha \beta} \partial_\mu B_\beta \right). \label{eq:EOM_1_p}
\end{align}
These equations describe the spin Hall effect of gravitational waves. The Hamiltonian, as well as Hamilton's equations have the same form as in the electromagnetic case presented in Ref. \cite[Eq. (4.15)-(4.17)]{GSHE2020}, except for the value of the constant $s$. The terms of $\mathcal{O}(\epsilon^1)$ are expressed in terms of the Berry connection, and they depend on the state of circular polarization through $s$. In the limit of infinitely-high frequencies, which corresponds to $\epsilon = 0$, we recover the geodesic equations, as in Eqs. \eqref{eq:EOM_0_x} and \eqref{eq:EOM_0_p}. 

As observed in Ref. \cite{GSHE2020}, the Hamiltonian \eqref{eq:H_canonical}, as well as the effective ray equations \eqref{eq:EOM_1_x} and \eqref{eq:EOM_1_p} are not independent of the choice of polarization vectors $m_\mu$ and $\bar{m}_\mu$. This is because the Berry connection $B_\mu$ is not invariant under spin rotations $m_\mu \mapsto e^{i \phi(x)} m_\mu$. Such transformations can be viewed as a change of gauge for the Berry connection. This is similar to the case of a charged particle moving in an electromagnetic field, and described by the minimally coupled Hamiltonian
\begin{equation}
    H = \frac{1}{2} g^{\mu \nu} (p_\mu - e A_\mu) (p_\nu - e A_\nu),
\end{equation}
which is not invariant under gauge transformations of the electromagnetic vector potential, $A_\mu \mapsto A_\mu + \nabla_\mu \xi$. Generally, this issue can be solved by introducing noncanonical coordinates, such that the connection one-form (e.g. the electromagnetic vector potential $A_\mu$ for the case of charged particles, or the Berry connection $B_\mu$ for the case of spinning particles) is eliminated from the Hamiltonian, and the ray equations are expressed in terms of the curvature two-form (e.g. the Faraday tensor $F_{\mu \nu} = 2 \nabla_{[\mu} A_{\nu]}$ for the case of charged particles, or the Berry curvature for the case of spinning particles). This procedure is discussed in Ref. \cite{Sternberg1977} for the case of a charged particle, and in Ref. \cite{Littlejohn1991} for Hamiltonians involving the Berry connection. Also, it is generally the case that the effective ray equations describing spin Hall effects in optics or condensed matter physics are usually expressed in terms of the Berry curvature \cite{SOI_review,Bliokh2008,Ruiz2015,SHE_QM2,Berry_CM1,Berry_CM2}.

Noncanonical coordinates for a Hamiltonian of the form given in Eq. \eqref{eq:H_canonical} were introduced in Ref. \cite{GSHE2020}, based on the general proposal of Littlejohn and Flynn \cite{Littlejohn1991}. The relation between canonical coordinates $(x^\mu, p_\mu)$ and noncanonical coordinates $(X^\mu, P_\mu)$ is
\begin{align} \label{eq:coord}
    X^\mu &= x^\mu + i \epsilon s \bar{m}^{\alpha} \vnabla{}^\mu {m}_\alpha, \\
    P_\mu &= p_\mu - i \epsilon s \bar{m}^{\alpha} \nabla_\mu {m}_\alpha. \label{eq:coord1}
\end{align}
The coordinate transformation is performed perturbatively with respect to $\epsilon$, and terms of $\mathcal{O}(\epsilon^2)$ are ignored. We refer the reader to Ref. \cite{GSHE2020} for the details of the calculations. In noncanonical coordinates $(X^\mu, P_\mu)$, the Hamiltonian is
\begin{equation}
    H(X, P) = \frac{1}{2}g^{\mu \nu}(X) P_\mu P_\nu,
\end{equation}
and the effective ray equations become
\begin{align}
    \begin{split}\dot{X}^\mu \, &= \, P^\mu + \epsilon s P^\nu \left( F_{p x} \right)\indices{_\nu^\mu} + \epsilon s \Gamma^\alpha_{\beta \nu} P_\alpha P^\beta \left( F_{p p}\right)^{\nu \mu} ,\end{split} \label{eq:Xdot}  \\
    \begin{split}\dot{P}_\mu \, &= \, \Gamma^\alpha_{\beta \mu} P_\alpha P^\beta - \epsilon s P^\nu \left( F_{x x} \right)_{\nu \mu} - \epsilon s \Gamma^\alpha_{\beta \nu} P_\alpha P^\beta \left( F_{x p} \right)\indices{^\nu_\mu} .\end{split} \label{eq:Pdot}
\end{align}
In the above equations, we have the components of the Berry curvature, defined as
\begin{equation} \label{eq:Berry_curvature}
\begin{split}
    &\begin{split} \left({F_{p p}}\right)^{\nu \mu} = i \Big( &\vnabla{}^\mu \bar{m}^\alpha  \vnabla{}^\nu m_\alpha - \vnabla{}^\nu \bar{m}^\alpha \vnabla{}^\mu m_\alpha + \bar{m}^\alpha \vnabla{}^{[\mu} \vnabla{}^{\nu]} m_\alpha - m_\alpha \vnabla{}^{[\mu} \vnabla{}^{\nu]} \bar{m}^\alpha \Big), \end{split} \\
    &\begin{split} \left({F_{xx}}\right)_{\nu \mu} = i \Big( &\nabla_\mu \bar{m}^\alpha \nabla_\nu m_\alpha - \nabla_\nu \bar{m}^\alpha \nabla_\mu m_\alpha + \bar{m}^\alpha \nabla_{[\mu} \nabla_{\nu]} m_\alpha - m_\alpha \nabla_{[\mu} \nabla_{\nu]} \bar{m}^\alpha \Big), \end{split} \\
    &\begin{split}\left({F_{p x}}\right)\indices{_\nu^\mu} &= -\left({F_{x p}}\right)\indices{^\mu_\nu} = i \left( \vnabla{}^\mu \bar{m}^\alpha \nabla_\nu m_\alpha - \nabla_\nu \bar{m}^\alpha \vnabla{}^\mu m_\alpha \right).\end{split}
\end{split}
\end{equation}
It can easily be verified that these equations are invariant under spin rotations $m_\mu \mapsto e^{i \phi(x)} m_\mu$. However, given a null covector $P_\mu$ the orthogonal plane spanned by $m_\mu$ and $\bar{m}_\mu$ is not uniquely fixed, since one can always perform transformations of the form $m_\mu \mapsto m_\mu + c P_\mu$. This orthogonal plane can only be fixed uniquely by introducing additional structure, such as a timelike vector $t^\mu$ or another null vector $n^\mu$, orthogonal to $m_\mu$ and $\bar{m}_\mu$. From a physical point of view, this means that the orthogonal plane spanned by $m_\mu$ and $\bar{m}_\mu$ can only be fixed with respect to a timelike observer with $4$-velocity $t^\mu$. 

As discussed in Ref. \cite{GSHE2020}, changing the vector field $t^\mu$, defining a family of observers, corresponds to a change of polarization vectors of the form $m_\mu \mapsto m_\mu + c P_\mu$. The effective ray equations \eqref{eq:Xdot} and \eqref{eq:Pdot} are not invariant under such transformations. This reflects the well-known fact that the position of a massless spinning particle cannot be defined independent of an observer. In particular, this can be viewed as a manifestation of the relativistic Hall effect \cite{Relativistic_Hall} and the Wigner translation for massless spinning particles \cite{Stone2015,DUVAL2015} (see also Ref. \cite{Herdeiro2012} for a similar discussion in the context of the Mathisson-Papapetrou-Dixon equations). It has been shown in Ref. \cite{GSHE2020} how Eqs. \eqref{eq:Xdot} and \eqref{eq:Pdot} incorporate these effects.

\section{Conclusion}
We have presented a covariant WKB analysis of gravitational waves, as described by the  linearized Einstein equations. By going beyond the standard geometrical optics approach, we obtained effective ray equations containing polarization-dependent terms and describing the spin Hall effect of gravitational waves propagating on arbitrary spacetimes. The effective ray equations have the same form as in the electromagnetic case discussed in Ref. \cite{GSHE2020}, the only difference being a factor of 2, representing the spin-2 nature of the gravitational field. Thus, considering electromagnetic and gravitational waves of the same frequency, the spin Hall effect is twice as large in the case of gravitational waves. 

In an ongoing work \cite{HarteOancea} (see also \cite{Oanceathesis}), the authors prove that the resulting equations can be cast in the form of the Mathisson-Papatreou-Dixon equations for massless particles, with the Corinaldesi-Papapetrou spin supplementary condition. The latter is a consequence of the derivation of the effective equations of motions. Furthermore, with \cite{GSHE2020}, it provides a first systematic covariant derivation of the equations of motions for massless spinning particles. 

The spin Hall effect of gravitational waves is expected to play an important role for gravitational waves of finite frequency. Hence, one important perspective is to understand the observable consequences of corrections to geometrical optics. Firstly, the corrections to geometrical optics should lead to measurable frequency-dependent corrections to gravitational lensing, as discussed in \cite{2020arXiv200812814M, PhysRevD.101.044041}. To calculate the effect, an analytic discussion of the effective equations of motions must be performed. Secondly, the effect measured is spin-dependent. The effective equations of motions should lead to different trajectories for electromagnetic and gravitational wave packets. This could lead to different arrival times. These aspects will be investigated in future works.

\section*{Acknowledgments}

M.A.O. is grateful for support by the International Max Planck Research School for Mathematical and Physical Aspects of Gravitation, Cosmology and Quantum Field Theory.

\appendix

\section{Properties of the symbol $D\indices{_{\alpha \beta}^{\gamma \delta}}$} \label{app:symbol}

The kernel of the symbol $D\indices{_{\alpha \beta}^{\gamma \delta}}$, considered as a endormorphism of the space of symmetric two-tensors, is calculated in this section. We first observe that, if $b_\delta$ is any covector, then 
\begin{equation}
D\indices{_{\alpha \beta}^{\gamma \delta}} k{}_{(\gamma} b{}_{\delta)} =0.
\end{equation}
The tensor $k{}_{(\gamma} b{}_{\delta)}$ is always in the kernel of $D\indices{_{\alpha \beta}^{\gamma \delta}}$.  More generally, if $S_{\gamma \delta}$ is a symmetric complex 2-tensor in the kernel of $D\indices{_{\alpha \beta}^{\gamma \delta}}$, then
\begin{align} 
 2 D\indices{_{\alpha \beta}^{\gamma \delta}}  S_{\gamma \delta} &= 
k_\alpha k_\beta S +g_{\alpha \beta} S_{\gamma \delta} k^\gamma k^\delta - k^\gamma S{}_{\gamma\alpha} k{}_{\beta }-  k^\gamma S{}_{\gamma\beta} k{}_{\alpha} \label{eq:kernelsymbol} \\
&= 0.  
\end{align}
We consider a Newman-Penrose tetrad $\{k_\alpha,n_\alpha,m_\alpha,\bar{m}_\alpha\}$ satisfying the orthogonality relations given in Eq. \eqref{eq:NPtetrad}. Considering symmetric tensor products of the Newman-Penrose tetrad elements, the only nontrivial contraction with the right-hand-side of Eq. \eqref{eq:kernelsymbol} are those with $m^\alpha \bar{m}^\beta$, $\bar{m}^\alpha {m}^\beta$,  $n^\alpha {m}^\beta$, $n^\alpha \bar{m}^\beta$,
\begin{equation}
    k^\gamma m^\beta S_{\gamma \beta} =k^\gamma \bar{m}^\beta S_{\gamma \beta} = k^\gamma k ^\beta S_{\gamma \beta}  =0,
\end{equation}
and $n^\alpha n^\beta$,
\begin{equation} \label{eq:trace_S}
S - 2 n^\alpha k^\beta S_{\alpha \beta} =0 = -2  S_{\alpha \beta} m^\alpha \bar{m} ^\beta. 
\end{equation} 
A similar argument can be made when $k^\mu$ is not null. Hence, we obtain the following lemma: 
\begin{lemma} \label{lem:kernelD}
When $k^\mu$ is a null vector, the kernel of the symbol $D\indices{_{\alpha \beta}^{\gamma \delta}}$ is the vector space of complex symmetric two-tensors generated by
\begin{gather}
    k_{\alpha}k_{\beta},  k{}_{(\alpha}n{}_{\beta)}, k{}_{(\alpha}m{}_{\beta)}, k{}_{(\alpha}\bar m{}_{\beta)},\label{eq:puregauge}\\
    m{}_{(\alpha}m{}_{\beta)},  \bar{m}{}_{(\alpha}\bar{m}{}_{\beta)}. \label{eq:polarization_A}
\end{gather}
When $k^\mu$ is not a null vector, the elements of the kernel of $D\indices{_{\alpha \beta}^{\gamma \delta}}$ are traceless symmetric two-tensors satisfying 
\begin{equation}
k^\alpha S_{\alpha\beta} = 0.
\end{equation}
\end{lemma}
Using Eq. \eqref{eq:trace_S}, one checks easily that, if $S_{\gamma \delta}$ is in the kernel of $D\indices{_{\alpha \beta}^{\gamma \delta}}$, then its trace-reverse $\check{S}_{\gamma \delta}$ satisfies
\begin{equation}
k^\alpha \check{S}_{\alpha\beta} = 0,   
\end{equation}
which is the form of the polarization tensor given in Eq. \eqref{eq:form of the polarization tensor}.

Finally, we observe that two-tensors generated by the elements of Eq. \eqref{eq:puregauge} are pure gauge. The Riemann curvature tensor of the particular perturbed metric tensor $\tilde{g}_{\alpha \beta} = g_{\alpha \beta} + \mathrm{Re} \left( k{}_{(\alpha} b_{ \beta )} e^{i S / \epsilon}\right)$, for an arbitrary $k_\alpha =\nabla_\alpha S$ and $b_\alpha$ complex covector, is given by, 
\begin{align}
\tilde R^\mu{}_{\nu \alpha \beta}  &    =R^\mu{}_{\nu \alpha \beta} +\nabla_\alpha \Gamma(h) ^\mu{}_{\nu \beta} -  \nabla_\beta \Gamma(h) ^\mu{}_{\nu \alpha}\\
\tilde R^\mu{}_{\nu \alpha \beta}  & =R^\mu{}_{\nu \alpha \beta}  + \mathcal{O}(\epsilon^{-1}), 
\end{align} 
instead of the expected
\begin{equation}
\tilde R^\mu{}_{\nu \alpha \beta}   =R^\mu{}_{\nu \alpha \beta}  + \mathcal{O}(\epsilon^{-2}).
\end{equation} 
Hence, a perturbation of the form $h_{\alpha \beta}= \mathrm{Re}\left( k{}_{(\alpha} b_{ \beta )} e^{i S / \epsilon} \right)$ is pure gauge at the lowest order in $\epsilon$.
\begin{lemma} The only non pure-gauge solutions of 
\begin{equation}
D\indices{_{\alpha \beta}^{\gamma \delta}}  S_{\gamma \delta} = 0
\end{equation}
are generated by 
\begin{equation}
m{}_{(\alpha}m{}_{\beta)},  \bar{m}{}_{(\alpha}\bar{m}{}_{\beta)}.
\end{equation}
\end{lemma}

\section{Derivation of the Lagrangian for linearized gravity}\label{app:linearization}

In this section, we consider the full metric $\tilde{g}_{\alpha \beta}$, written as a sum of a background metric $g_{\alpha \beta}$, and a small perturbation metric $h_{\alpha \beta}$:
\begin{equation}
    \tilde{g}_{\alpha \beta} = g_{\alpha \beta} + h_{\alpha \beta}.
\end{equation}
Recall that, with our conventions we have
\begin{equation}
    \tilde{g}^{\alpha \beta} = g^{\alpha \beta} - h^{\alpha \beta} +\mathcal{O}(|h|^2).
\end{equation}
The Einstein-Hilbert action is for the full metric $\tilde{g}_{\alpha \beta}$ is
\begin{equation}
\int_M \mathrm{d}^4 x \, \sqrt{\tilde{g}} \,  \widetilde{R}.
\end{equation}
As always, the linearization of the determinant of the metric tensor leads to 
\begin{equation}
\sqrt{\tilde{g}} = \sqrt{{g}} \left( 1 + \dfrac12 g^{\alpha \beta}h_{\alpha \beta}\right) + \mathcal{O}(|h|^2).
\end{equation}
We introduce the notation
\begin{equation}
\begin{split}   
\tilde \Gamma^\alpha_{\beta \gamma} &= \Gamma^\alpha_{\beta \gamma} + \Upsilon^\alpha_{\beta \gamma},\\
\Upsilon^\alpha_{\beta \gamma} & = \dfrac12 g^{\alpha \sigma }\left(-\nabla _\sigma h_{\beta \gamma} +\nabla_\beta  h_{\sigma\gamma}   +\nabla_\gamma h_{\beta\sigma } \right) + \mathcal{O}(|h|^2),
\end{split}
\end{equation}
where the Christoffel symbols $\Gamma^\alpha_{\beta \gamma}$ and the covariant derivative $\nabla_\alpha$ are defined with respect to the background metric $g_{\alpha \beta}$. As the difference between two the Christofell symbols of two metrics, $\Upsilon^\alpha_{\beta \gamma}$ is a tensor. Now, we expand the Riemann curvature tensor of $\tilde{g}_{\alpha \beta}$,
\begin{equation}
 \begin{split}   
\tilde R^\mu{}_{\nu \alpha \beta}  & = R^\mu{}_{\nu \alpha \beta} +  \tilde{\nabla}_\alpha \Upsilon^\mu_{\nu \beta} - \tilde{\nabla}_\beta \Upsilon^\mu_{\nu \alpha} + 2 \left( \Upsilon^{\mu}_{\sigma\beta} \Upsilon^{\sigma}_{\nu\alpha}  - \Upsilon^{\mu}_{\sigma\alpha} \Upsilon^{\sigma}_{\nu\beta}   \right),
\end{split}
\end{equation}
where $\tilde{\nabla}_\alpha$ is the covariant derivative defined with respect to $\tilde{g}_{\alpha \beta}$. We contract in $\mu$ and $\alpha$ to get the Ricci curvature, and with inverse metric tensor $\tilde{g}^{\nu \beta}$ to get the scalar curvature:
\begin{align}
\widetilde{R}_{\nu \beta} & = \tilde R^\mu{}_{\nu \mu \beta} = {R}_{\nu \beta} + \tilde{\nabla}_\mu \Upsilon^\mu_{\nu \beta} - \tilde{\nabla}_\beta \Upsilon^\mu_{\nu \mu}    + 2 \left( \Upsilon^{\mu}_{\sigma\beta} \Upsilon^{\sigma}_{\nu\mu}  - \Upsilon^{\mu}_{\sigma\mu} \Upsilon^{\sigma}_{\nu\beta}   \right),     \\
\widetilde{R} & =\tilde{g}^{\nu \beta}  \widetilde{R}_{\nu \beta} = \tilde{g}^{\nu \beta}  {R}_{\nu \beta}   + \tilde{g}^{\nu \beta}  \left( \tilde{\nabla}_\mu \Upsilon^\mu_{\nu \beta} - \tilde{\nabla}_\beta \Upsilon^\mu_{\nu \mu} \right)   + 2 \tilde{g}^{\nu \beta}  \left( \Upsilon^{\mu}_{\sigma\beta} \Upsilon^{\sigma}_{\nu\mu}  - \Upsilon^{\mu}_{\sigma\mu} \Upsilon^{\sigma}_{\nu\beta}   \right).    
\end{align}
We consider now the Einstein-Hilbert action for the metric $\tilde{g}_{\alpha \beta}$:
\begin{equation}
\begin{split}
\int_M \mathrm{d}^4 x \, \sqrt{\tilde{g}} \,  \widetilde{R} &=
  \int_M \mathrm{d}^4 x \, \sqrt{\tilde{g}} \,  \tilde{g}^{\nu \beta}  \left[ {R}_{\nu \beta}  + 2   \left( \Upsilon^{\mu}_{\sigma\beta} \Upsilon^{\sigma}_{\nu\mu}  - \Upsilon^{\mu}_{\sigma\mu} \Upsilon^{\sigma}_{\nu\beta}   \right) \right] \\ 
 &\quad + \int_M \mathrm{d}^4 x \, \sqrt{\tilde{g}} \,   \tilde{g}^{\nu \beta}  \left( \tilde{\nabla}_\mu \Upsilon^\mu_{\nu \beta} - \tilde{\nabla}_\beta \Upsilon^\mu_{\nu \mu} \right).
    \end{split}
\end{equation}
In the above equation, the term on the second line is a boundary term, which we drop. In the first line, the second term can be rewritten as
\begin{align}
 \tilde{g}^{\nu \beta}  \left( \Upsilon^{\mu}_{\sigma\beta} \Upsilon^{\sigma}_{\nu\mu}  - \Upsilon^{\mu}_{\sigma\mu} \Upsilon^{\sigma}_{\nu\beta}   \right)  \sqrt{{\tilde g}}  &=  {g}^{\nu \beta}  \left( \Upsilon^{\mu}_{\sigma\beta} \Upsilon^{\sigma}_{\nu\mu}  - \Upsilon^{\mu}_{\sigma\mu} \Upsilon^{\sigma}_{\nu\beta}   \right)  \sqrt{{g}}   + \mathcal{O}(|h|^3).
\end{align}
Using the expansion of the determinant of the metric tensor, we obtain 
\begin{equation}
    \tilde{g}^{\nu \beta}   \sqrt{{\tilde g}}   =  \sqrt{{ g}} \left( g^{\nu \beta} - h^{\nu\beta} +\frac12  h g^{\nu\beta}  \right) +\mathcal{O}(|h|^2). 
\end{equation}
The expansion of the Einstein-Hilbert action, neglecting terms of order $3$ in $h_{\alpha \beta}$, takes the preliminary form
\begin{equation}
\begin{split}
    \int_M \mathrm{d}^4 x \, \sqrt{\tilde{g}} \,  \widetilde{R} &= \int_M \mathrm{d}^4 x \, \sqrt{g} \,  {R} -  \int_M \mathrm{d}^4 x \, \sqrt{g} \, \left( R_{\nu \beta}  -  \dfrac12 R  g_{\nu \beta}\right) h^{\nu \beta} \\
     &\quad  +  \int_M \mathrm{d}^4 x \, \sqrt{g} \,  R_{\mu\nu } \mathcal{V}^{\mu\nu} \\
     &\quad +  \int_M \mathrm{d}^4 x \, \sqrt{g} \,  2 g^{\nu \beta}  \left( \Upsilon^{\mu}_{\sigma\beta} \Upsilon^{\sigma}_{\nu\mu}  - \Upsilon^{\mu}_{\sigma\mu} \Upsilon^{\sigma}_{\nu\beta}   \right)   + \mathcal{O}(|h|^3).
\end{split}
\end{equation}
where $ \mathcal{V}^{\mu\nu}   = \mathcal{O}(|h|^2)  $.  Since we assume that the background metric $g_{\alpha \beta}$ satisfies the Einstein field equations in vacuum with no cosmological constant, we have
\begin{equation}
    \int_M \mathrm{d}^4 x \, \sqrt{\tilde{g}} \,  \widetilde{R} =  \int_M \mathrm{d}^4 x \, \sqrt{g} \,   g^{\nu \beta}  \left( \Upsilon^{\mu}_{\sigma\beta} \Upsilon^{\sigma}_{\nu\mu}  - \Upsilon^{\mu}_{\sigma\mu} \Upsilon^{\sigma}_{\nu\beta} \right)  + \mathcal{O}(|h|^3).
\end{equation}
Using the definition of $\Upsilon^\alpha_{\beta \gamma}$, we can calculate
\begin{align}
    g^{\nu \beta}  \Upsilon^{\mu}_{\sigma\beta} \Upsilon^{\sigma}_{\nu\mu}  &= \frac14\left( 2\nabla^{\sigma} h^{\mu\nu}\nabla_{\mu}h_{\sigma\nu}  - \nabla^{\sigma} h^{\mu\nu}\nabla_{\sigma} h_{\mu\nu} \right)+  \mathcal{O}(|h|^3) \\ 
    g^{\nu \beta}    \Upsilon^{\mu}_{\sigma\mu} \Upsilon^{\sigma}_{\nu\beta}  &=   \frac14 \left( -\nabla_\sigma h \nabla^\sigma h  + 2\nabla^\sigma h \nabla^{\mu} h_{\sigma\mu}  \right)+  \mathcal{O}(|h|^3). 
\end{align}
The linearized Einstein-Hilbert action, neglecting terms of order $3$ in $h_{\alpha \beta}$, is given by
\begin{equation}
\begin{split}
    \int_M \mathrm{d}^4 x \, \sqrt{\tilde{g}} \,  \widetilde{R} &=  \int_M \mathrm{d}^4 x \, \sqrt{g} \, L + \mathcal{O}(|h|^3),
\end{split}
\end{equation} 
where $L$ is the Lagrangian for linearized gravity, defined as
\begin{equation}
  L =   \nabla^{\sigma} h^{\mu\nu}\nabla_{\mu}h_{\sigma\nu}  - \frac12 \nabla^{\sigma} h^{\mu\nu}\nabla_{\sigma} h_{\mu\nu}   + \frac12 \nabla_\sigma h \nabla^\sigma h  - \nabla^\sigma h \nabla^{\mu} h_{\sigma\mu} ,
\end{equation}
which agrees with the Lagrangian obtained in Ref. \cite[p. 55]{1971bicak}. Integrating by parts and neglecting boundary terms, we obtain
\begin{equation}
\begin{split}
        \int_M \widetilde{R} \sqrt{\tilde{g}} dx  = \int_M \mathrm{d}^4 x \, \sqrt{g} \,  h^{\alpha \beta} \hat{D}\indices{_{\alpha \beta}^{\gamma \delta}} h_{\gamma \delta},   
\end{split}
\end{equation}
where $\hat{D}\indices{_{\alpha \beta}^{\gamma \delta}}$ is defined as
\begin{equation}
\hat{D}\indices{_{\alpha \beta}^{\gamma \delta}} = \frac{1}{2} \left( \delta_\alpha^\gamma \delta_\beta^\delta \nabla_\mu \nabla^\mu  - g_{\alpha \beta} g^{\gamma \delta}\nabla_\mu \nabla^\mu  + g^{\gamma \delta}\nabla_\alpha \nabla_\beta  + g_{\alpha \beta} \nabla^\gamma \nabla^\delta - \delta_\beta^\delta \nabla^\gamma\nabla_\alpha  - \delta_\alpha^\delta\nabla^\gamma \nabla_\beta  \right).
\end{equation}

\section{Lorenz gauge} \label{app:gauge}
 
\subsection{Linearization of the wave gauge}
   
We start by the standard calculation of the linearization of the wave gauge. Consider a chart $(U, x^\alpha)$, and assume that this chart is harmonic for the metric ${\tilde g}_{\alpha \beta}$. That is
\begin{equation}
      \tilde{ g}^{\alpha\beta}\tilde{\nabla}_{\alpha}\tilde{\nabla}_\beta x^\delta =F^\delta,
\end{equation}
where the $F^\delta$ are unknown functions to be chosen wisely. We expand this to get
\begin{equation}
\begin{split}
       F^\delta &=  \tilde{g}^{\alpha\beta}\tilde{\nabla}_{\alpha}\tilde{\nabla}_\beta x^\delta\\
       &=\tilde{g}^{\alpha\beta}\partial_{x^\alpha} \partial_{x^\beta} x^\delta  +  \tilde{g}^{\alpha\beta} \tilde{\Gamma}_{\alpha \beta} ^\mu \partial_{x^\mu} x^\delta.\\
      &=   \tilde{g}^{\alpha\beta} \tilde{\Gamma}_{\alpha \beta}^\delta
          \end{split}
\end{equation}
Using
\begin{align}
    \tilde \Gamma^\alpha_{\beta \gamma} &= \Gamma^\alpha_{\beta \gamma} + \dfrac12 g^{\alpha \sigma }\left(-\nabla _\sigma h_{\beta \gamma} +\nabla_\beta  h_{\sigma\gamma}   +\nabla_\gamma h_{\beta\sigma } \right), 
\end{align}
and neglecting the quadratic terms in $h_{\alpha \beta}$, we obtain
\begin{align}
   \underbrace{{g}^{\beta \gamma}\Gamma^\alpha_{\beta \gamma}}_{\text{order 0 in $h_{\alpha \beta}$}}
   -    \underbrace{ h^{\beta \gamma}\Gamma^\alpha_{\beta \gamma} +  \dfrac{1}{2} \left( 2\nabla_\beta h^{\beta \alpha} -\nabla^\alpha h\right)}_{\text{order 1 in $h_{\alpha \beta}$}}  &=F^\delta \\
   \nabla_\beta \left(  h^{\beta \alpha}  - \dfrac12 h g^{\beta \alpha} \right) &= F^\delta - \tilde{g}^{\beta \gamma}\Gamma^\alpha_{\beta \gamma}.
\end{align}
For a general background metric, we choose
\begin{equation}
F^\delta = \tilde{g}^{\beta \gamma}\Gamma^\alpha_{\beta \gamma}
\end{equation}
in order to obtain the Lorenz gauge condition
\begin{equation}
   \nabla_\beta \left(  h^{\beta \alpha}  - \dfrac12 h g^{\beta \alpha} \right)= 0.
\end{equation}
When $g_{\alpha \beta}$ is the Minkowski metric, and $(U, x^\alpha)$ the Cartesian chart on $\mathbb{R}^4$, then $F^\delta$ can be chosen equal to $0$.

\subsection{Propagation of the gauge} 

In that section, we check that the gauge condition is conserved by the equation for linearized gravity. This is a linearization of the procedure described in Ref. \cite[Chapter 14.2]{MR2527641}. We introduce 
\begin{equation}
    \mathcal{G}_\mu = \nabla^\alpha h_{\alpha \mu} - \dfrac{1}{2} \nabla_{\mu } h.
 \end{equation}
Recall that $\breve{h}_{\alpha \beta}$ is the trace-reversed of $h_{\alpha \beta}$. Observe that Eq. \eqref{eq:MTW_Wave eqn} can be rewritten as
\begin{equation} \label{C11}
-\nabla_\alpha \nabla^\alpha \breve{h}_{\mu \nu} -g_{\mu \nu} \nabla_\alpha \nabla_\beta  \breve{h}^{\alpha \beta}  + \nabla^\alpha \nabla_\mu \breve{h}_{\alpha \nu} + \nabla^\alpha \nabla_\nu \breve{h}_{\alpha \mu } = 0.
\end{equation}
By commuting the covariant derivatives as
\begin{align} 
\nabla^\alpha \nabla_\mu \breve{h}_{\alpha \nu} & = \nabla_\mu  \nabla^\alpha \breve{h}_{\alpha \nu} +R_{\nu\alpha\sigma  \mu  } \breve{h}^{\sigma \alpha},
\end{align}
and using the fact that $g_{\alpha \beta}$ is Ricci flat, we obtain
\begin{equation} \label{eq:gaugestep2}
\nabla_\alpha \nabla^\alpha \breve{h}_{\mu \nu} - 2 R_{\nu\alpha\sigma  \mu  }  \breve{h}^{\sigma \alpha}  =  \nabla_{\mu}    \mathcal{G}_\nu +\nabla_{\nu}    \mathcal{G}_\mu - g_{\mu \nu} \nabla^{\alpha}    \mathcal{G}_\alpha  .
\end{equation}
Taking the divergence of the right-hand side of the previous equations, we obtain
\begin{equation} 
\nabla_\alpha \nabla^\alpha  \mathcal{G}_\mu + R_{\mu \alpha}  \mathcal{G}^\alpha = \nabla^\mu \left(  \nabla_\alpha \nabla^\alpha \breve{h}_{\mu \nu}  - 2 R_{\nu\alpha\sigma  \mu  }  \breve{h}^{\sigma \alpha} \right).
\end{equation}
Hence, if we consider a solution of the reduced equation 
\begin{equation} 
\nabla_\alpha \nabla^\alpha \breve{h}_{\mu \nu}  - 2 R_{\nu\alpha\sigma  \mu  }  \breve{h}^{\sigma \alpha}  = 0
\end{equation}
and we assume that, initially, 
\begin{equation} 
\mathcal{G}_\mu  =0 \quad \text{ and } \quad \nabla_\nu \mathcal{G}_\mu  =0,  
\end{equation}
then ${h}_{\alpha \beta}$ solves Eq. \eqref{eq:MTW_Wave eqn} in the Lorenz gauge. Furthermore, the trace of $h_{\alpha \beta}$ satisfies the decoupled equation equation 
\begin{equation} 
\nabla^\mu \nabla_\mu h = 0.
\end{equation}
%

\bibliographystyle{abbrv}
\bibliography{references}

\end{document}